\documentclass[12pt]{article}
\usepackage{graphics, color}
\usepackage{graphicx}
\usepackage{amssymb, amsmath, tensor}
\pdfoutput=1

\newcommand{\sect}[1]{\section{#1}\setcounter{equation}{0}}

\def\gsim{\, \rlap{$>$}{\lower 1.1ex\hbox{$\sim$}}\,}
\def\lsim{\, \rlap{$<$}{\lower 1.1ex\hbox{$\sim$}}\,}

\begin{document}


\begin{titlepage}
\bigskip
\bigskip\bigskip\bigskip
\centerline{\Large Holographic and Wilsonian Renormalization Groups}
\bigskip\bigskip\bigskip
\bigskip\bigskip\bigskip

 \centerline{{\bf Idse Heemskerk}\footnote{\tt idse@physics.ucsb.edu}}
\medskip
\centerline{\em Department of Physics}
\centerline{\em University of California}
\centerline{\em Santa Barbara, CA 93106}
\bigskip
 \centerline{{\bf Joseph Polchinski}\footnote{\tt joep@kitp.ucsb.edu}}
\medskip
\centerline{\em Kavli Institute for Theoretical Physics}
\centerline{\em University of California}
\centerline{\em Santa Barbara, CA 93106-4030}\bigskip
\bigskip
\bigskip\bigskip


\begin{abstract}
We develop parallels between the holographic renormalization group in the bulk and the Wilsonian renormalization group in the dual field theory. Our philosophy differs from most previous work on the holographic RG; the most notable feature is the key role of multi-trace operators.  We work out the forms of various single- and double-trace flows.  The key question, `what cutoff on the field theory corresponds to a radial cutoff in the bulk?' is left unanswered, but by sharpening the analogy between the two sides we identify possible directions.
\end{abstract}
\end{titlepage}
\baselineskip = 16pt

\tableofcontents

\section{Introduction}

In a sense, AdS/CFT duality reduces the construction of quantum gravity to a solved problem.  Gravity is written in terms of a nongravitational quantum field theory~\cite{Maldacena:1997re,Gubser:1998bc,Witten:1998qj}, and the technology of Wilson~\cite{Wilson:1993dy}
 can then be applied.  In its current form, this duality reports only the measurements made by an external observer studying gravity confined to an anti-de Sitter box.  Still, this allows one to address many of the conceptual questions of quantum gravity, most notably the purity of black hole Hawking radiation.  This implies in turn an extreme nonlocality of quantum gravity~\cite{Susskind:1993if}.     Formulating a theory in which locality has been so radically abandoned is a great challenge.  Fortunately, we have the example of AdS/CFT duality, and so we should understand this example as fully as possible.
 
In retrospect, what AdS/CFT describes are the simplest and most sharply defined observables possible in a theory of quantum gravity.  One needs go beyond these, to describe local observers within AdS space, and to understand gravity outside the AdS box.   In this paper we study a framework suggested by the discussion above, in which we pull the Wilson renormalization group back through the duality, where it takes a holographic form.  Of course there is a wide literature that falls under the broad heading of {\it holographic renormalization group (HRG)}.   However, the parallels between the Wilson and holographic perspectives have not been fully developed, and there are some simple observations and exercises that we would like to add, although we will also leave many questions unanswered for now.

Many of the points that we make have already been made in various forms in the literature, but it is useful to bring together here a coherent point of view.
Although we have emphasized the use of the duality to understand quantum gravity, clarifying the relation between the RG's should also be useful in the applications to strongly coupled field theories.  Indeed, there is interesting overlap between some recent work in this area and ours.

Wilson separates the path integral into high- and low-energy modes, and integrates out the former,
\begin{equation}
Z = \int {\cal D}M_{k \delta < 1}\,  {\cal D}M_{k \delta > 1}\, e^{-S} = \int {\cal D}M_{k \delta < 1}\,
e^{-S(\delta)} \,,
\label{wilsplit}
\end{equation}
where we use $M$ for generic boundary fields.  We are being impressionistic here: the precise form of the cutoff at scale $\delta$ is not specified.
The progressive integration of momentum shells is then described by a differential equation, and by integrating out to $\delta = \infty$ one obtains the full path integral.  The intermediate stages are quite complicated: $S(\delta)$ is necessarily a quasilocal function of length scale $\delta $, and so depends on an infinite number of parameters, but the flow is strongly convergent in almost all directions.  In essence, this flow supplies the deltas and epsilons of path integration, reducing the construction of QFT to dimensional analysis.  In the spirit of full disclosure we should note that it is difficult to make the mode separation while preserving explicitly important structures such as gauge invariance, supersymmetry, and duality.  Ultimately this may point to the need for a new way to think about QFT, but for now this is the only tool of broad applicability, enabling one to reduce arbitrary calculations to algorithms.

In AdS/CFT duality, the radial coordinate $z$ of the AdS bulk emerges from the energy scale of the boundary field theory~\cite{Maldacena:1997re,Gubser:1998bc,Witten:1998qj,Susskind:1998dq}.  This suggests a correspondence between radial evolution in the bulk and Wilson renormalization group flow in the field theory.  In Sec.~2 we develop this idea.  Our approach is to postulate a mapping between the radial cutoff on the bulk, and a Wilsonian cutoff on the field theory.  To make the form of this mapping precise is the deep and difficult problem that we would like to solve in the future, but for the present we are interested in the consequences of this postulate for the structures on the two sides.  Two somewhat unexpected results are the central role played by multi-trace operators, and the need to carry out the gravitational path integral in gauge-fixed rather than Wheeler-DeWitt form.  In Sec.~3 we work out some examples, in particular the flows induced by relevant single- and double-trace operators, and also flows involving bulk gauge fields and the metric.  In Sec.~4 we discuss  various truncations of the full Wilsonian flow and relations with the earlier literature.  In Sec.~5 we discuss the picture from the boundary field theory point of view, and are led to a possible, but rather unusual, regulator.  In Sec.~6 we discuss future directions and intended applications.

\sect{General structure}

In the bulk, our approach requires that we write the theory as a path integral over local fields.  Already this leads to some tension, as the Wilsonian framework requires that we keep all operators in the theory, including those dual to stringy states, and so it is not clear whether such a local representation is possible.  We will gloss over this point for now, and return to it in Sec.~5.

\subsection{Fixed background}

An asymptotically $AdS_{d+1}$ metric can be written
\begin{equation} \label{adsmet}
ds^2 = n^2(z,x) dz^2 + h_{\mu\nu}(z,x) dx^\mu dx^\nu \,, \quad \lim_{z\to0} n(z,x) = \frac{L}{z}\,, \quad \lim_{z\to 0} h_{\mu\nu}(z,x) = \frac{L}{z}\delta_{\mu\nu} \,.
\end{equation}
We use a lower-case $n$ for the lapse to distinguish it from the number of colors.
Some flows of interest involve large deformations of the spacetime, and so we must be must deal with the dynamics of the metric.  In order to get oriented, however, we will first consider flows for which the backreaction is small, and so take the path integral in the fixed metric~(\ref{adsmet}), focussing for simplicity on scalar fields.  Separate the path integration over the fields $\phi^i(z,x)$ into three parts, $z > \ell$, $z < \ell$, and $z=\ell$:
\begin{eqnarray}
Z &=& \int {\cal D}\phi\, e^{-\kappa^{-2}{\cal S}}  \nonumber\\
&=& \int {\cal D}\phi|^{\vphantom j}_{z > \ell}\, {\cal D}\tilde\phi \, {\cal D}\phi|^{\vphantom j}_{z < \ell}\, e^{-\kappa^{-2}{\cal S}|^{\vphantom j}_{z > \ell} - \kappa^{-2}{\cal S}|^{\vphantom j}_{z < \ell}}  \nonumber\\
&=& \int {\cal D}\tilde\phi \,  \Psi_{\rm IR}(\ell,\tilde\phi)  \Psi_{\rm UV}(\ell,\tilde\phi) \,.
\label{holosplit}
\end{eqnarray}
Here $\tilde\phi^i(x) = \phi^i(\ell,x)$, and $\kappa^{-2}{\cal S}$ denotes the bulk action, while $S$ and also $\kappa^{-2}s$ will be used for QFT actions. The dependence of the full amplitude $Z$ and of the UV piece $\Psi_{\rm UV}$  on the choice of AdS boundary conditions, including any sources, is left implicit. We keep track of $\kappa^2 \sim G_N$ (which for $AdS^5$ is $\sim L^3/N^2$) in anticipation of the classical limit, and we assume that $\mathcal{S}$ is of supergravity type, by which we mean that with canonical normalization it is independent of $\kappa$. 

This radial separation resembles a Wilsonian treatment of the dual field theory. In particular, it is natural to interpret the IR factor $\Psi_{\rm IR}(\ell,\tilde\phi)$ as the path integral in the dual QFT with a UV cutoff~\cite{Susskind:1998dq}, on a length scale $\delta$ to be determined shortly, 
\begin{equation}
\Psi_{\rm IR}(\ell,\tilde\phi) = \int {\cal D}M|^{\vphantom{j}}_{k \delta < 1}\, \exp \left\{-S_0 + \frac{1}{\kappa^2}\int d^dx \,\tilde\phi^i(x) {\cal O}_i(x) \right\} \,. \label{ir}
\end{equation}
Here ${\cal O}_i(x)$ are a complete set of local single trace operators built from the matrix fields $M$ and their derivatives.  They are normalized
\begin{equation}
 {\cal O}_i \sim \frac{1}{N} \mathrm{Tr} \left(\ \ldots \ \right)  \,,
 \end{equation}
 where the ellipsis represents a product of the matrix fields and their derivatives.
Thus $\tilde \phi^i$ and ${\cal O}_i$ are held fixed in the large-$N$ limit, and $\langle {\cal O}^i\rangle \sim N^0$.  It would be more standard to have a factor of $N^2$ in the exponent rather than $\kappa^{-2}$, which is of the same order in $N$, but this makes subsequent equations simpler.  The action $S_0$ is included for generality, though one might absorb it into the definition of the measure.  The linearity of the action in $\tilde\phi$ can be arranged by a bulk field redefinition.

{\it We take Eq.~(\ref{ir}) as a postulate,} and explore its consequences.  Of course we would like to be explicit about the precise form of the cutoff $\delta$ and action $S_0$.  This is the hard question, whose solution would allow us to discuss local physics in the bulk.  For now, though, we will simply assume this and see what it implies for the rest of the RG structure.  

Similarly, it is tempting to identify the UV factor $\Psi_{\rm UV}(\ell,\tilde\phi)$ with a Wilsonian action, integrating out the fields above the cutoff scale.  This action is not local, because propagating fields have been integrated out, but will fall exponentially beyond some length $\delta$ because the fields are confined to a finite range $0 < z < \ell$, and so there is a gap for propagation in the $x^\mu$ directions. This locality property has been discussed previously in Ref.~\cite{Mansfield:1999kk}.

The scale $\delta$ is estimated in Ref.~\cite{Susskind:1998dq} in terms of a geodesic connecting two boundary points and hanging down to radius $z$.  The same order of magnitude is given by the extent in $x^\mu$ of the lightcone of a boundary point when it touches $z=\ell$ \cite{Bousso:2009dm}, as illustrated in Fig.~1.
\begin{figure}
\begin{center}
\includegraphics[scale=.70]{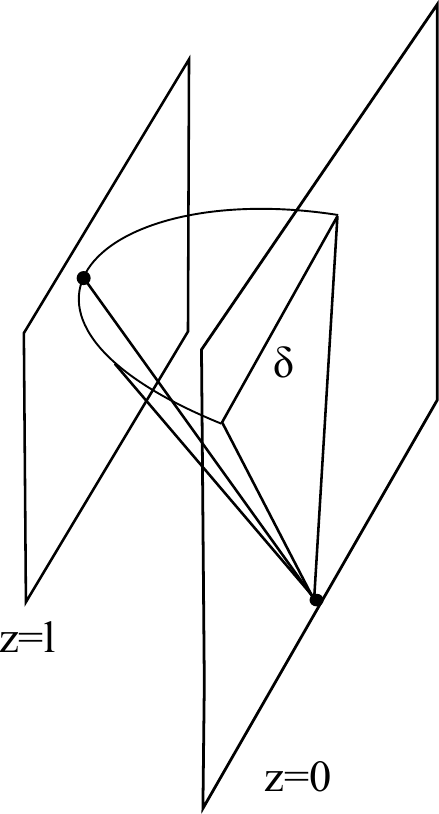} 
\caption{The scale $\delta$ on which the Wilsonian action is localized.}
\end{center}
\label{deltafig}
\end{figure}
For a Poincar\'e invariant metric $h_{\mu\nu}=e^{2A(z)}\delta_{\mu\nu}$ this is given by
\begin{equation} 
\delta \approx \int^{\ell}_0 dz\,n(z) e^{-A(z)}  = \int^{s(\ell)}_{-\infty} ds \,e^{-A(z(s))} \approx -\frac{e^{-A(\ell)}}{\partial_s A(\ell)}  \,.
\end{equation}
Here $s$ is the proper distance, and we assume that, as in AdS, $e^{-A(z)}$ grows rapidly toward the IR and so the integral can be estimated by the behavior near the upper limit.  For pure AdS space one simply has 
\begin{equation}
s= L \ln (z/L)\,, \quad A(z) = -s/L\,, \quad 
\delta(\ell) = \ell\,.
\end{equation}  
More generally,
\begin{equation}
e^{A(\ell)}\delta \sim -1/\partial_s A(\ell) \,,\label{cutoff}
\end{equation}
meaning that  the proper length corresponding to the cutoff is of the order of the effective AdS radius at the given scale.  This is also consistent with one degree of freedom per Planck area~\cite{Susskind:1998dq}, $N^2 \sim L^{d-1}/L_{\rm P}^{d-1}$ in terms of the $d+1$ dimensional Planck length.  The expansion in local terms thus involves an infinite number of parameters, as in the Wilsonian action.

Finally, in comparing the holographic and Wilsonian structures~(\ref{holosplit}) and (\ref{wilsplit}), what is the role of the functional integral over $\tilde\phi^i$?  It appears that rather than a definite RG flow we have instead some weighted average over effective couplings. To get some insight, consider a single scalar, suppose that the UV factor is a local gaussian\begin{equation}
\Psi_{\rm UV}(\ell,\tilde\phi)= \exp\left\{ -\frac{1}{2h\kappa^2} \int d^d x\, (\tilde\phi(x) + g(x))^2 \right\} \,.
\end{equation}
Using the postulate~(\ref{ir}) and carrying out the integral over $\tilde\phi$ gives
\begin{equation}
Z \propto  \int {\cal D}M|^{\vphantom{j}}_{k \delta < 1}\, \exp \left\{- S_0 -\frac{1}{\kappa^2} \int d^dx \left( g(x) {\cal O}(x) - \frac{h}{2} {\cal O}(x)^2 \right) \right\} \,.   \label{gauss}
\end{equation}
Thus we do obtain a definite effective action, but one that has double-trace as well as single-trace terms.\footnote{The double-trace terms enter with a negative sign, but if the original theory is stable it must be that the total action is stable.} More generally, we see that the Wilsonian action is not $\Psi_{\rm UV}(\ell,\phi)$ directly, but rather an integral transform of it,
\begin{equation}
\exp\left(-\kappa^{-2} s(\delta) \right) =\int {\cal D}\tilde\phi\, \exp \left\{\frac{1}{\kappa^{2}}\int d^dx \, \tilde\phi^i(x) {\cal O}_i(x) \right\} \Psi_{\rm UV}(\delta,\tilde\phi) \,. \label{wilson}
\end{equation}
The action $S(\delta) = \kappa^{-2} s(\delta)$ has general multi-trace terms, and is localized on the scale $\delta$.

That renormalization group flow generically leads to multi-trace operators, and moreover that these contribute to connected correlators even in the planar limit, has been discussed in a number of situations.  For example, in nonsupersymmetric orbifolds of the ${\cal N}=4$ theory this was observed in Refs.~\cite{Tseytlin:1999ii}, while in the Wilson RG context it was noted in Refs.~\cite{Li:2000ec,Petkou:2002bb};  we will review the latter work in Sec.~5.  Nevertheless, it comes as a bit of a surprise, because most of the HRG literature focuses on flows within the space of single-trace couplings, and because of the apparently misplaced prejudice that single-trace actions are the norm in taking large-$N$ limits, with multi-trace actions arising in exotic applications such as Refs.~\cite{Aharony:2001pa,Witten:2001ua,Hertog:2004rz}.

The total amplitude is independent of the radius at which the division is made,
\begin{equation}
0 = \frac{d}{d\ell} Z = \frac{d}{d\ell} \left\langle e^{-\kappa^{-2} s(\delta) }\right\rangle_ \delta \,,\label{invar}
\end{equation}
where the angle brackets denote the matrix path integral~(\ref{ir}), and the derivative in the last form acts on both the cutoff and the action.  The `amplitudes' $\Psi_{\rm IR}, \Psi_{\rm UV}$ evolve via radial Schr\"odinger equations~\cite{Mansfield:1999kk}
\begin{eqnarray}
\kappa^{2}\partial_\ell \Psi_{\rm IR}(\ell,\tilde\phi) &=& H(\tilde\phi, \tilde\pi) \Psi_{\rm IR}(\ell,\tilde\phi) \,, \nonumber\\
\kappa^{2}\partial_\ell \Psi_{\rm UV}(\ell,\tilde\phi) &=& -H(\tilde\phi,  \tilde\pi) \Psi_{\rm UV}(\ell,\tilde\phi) \,,
\label{evolve}
\end{eqnarray}
where $\tilde\pi = -i \kappa^2 \delta/\delta\tilde\phi$.
For example, for 
\begin{equation}
{\cal S} = \int d^dx\,dz\, \sqrt g \left(\frac{1}{2} \partial_M \phi \partial^M \phi + V(\phi) \right)\,,
\end{equation}
\begin{equation}
 H(\tilde\phi, \tilde\pi) = \int d^dx \,N \left( \frac{1}{2\sqrt{h}} \tilde\pi^2 + \frac{1}{2}\sqrt{h}h^{\mu\nu}\partial_\mu \tilde\phi \partial_\nu \tilde\phi + \sqrt{h}V(\tilde\phi)\right) \,.
\end{equation}

For the Wilsonian action~(\ref{wilson}), the evolution equation becomes
\begin{equation}
\kappa^2 \partial_\ell e^{-\kappa^{-2} s(\delta)} =
\kappa^2 \partial_\ell \delta \,\partial_\delta e^{-\kappa^{-2} s(\delta)} =
 -H(\kappa^2 \delta/\delta{\cal O}, i{\cal O}) e^{-\kappa^{-2} s(\delta)} \,. \label{wilrg}
\end{equation}
This is our main result, an equation that should be compared in form to that found in the QFT.
In the classical limit $\kappa^2\to 0$, corresponding to the planar limit of the QFT, this reduces to the Hamilton-Jacobi (HJ) equation
\begin{equation}
\partial_\ell s(\delta) = H(-\delta s(\delta) /\delta{\cal O},i{\cal O})\,. 
\label{hjs}
\end{equation}

Let ${\cal A}_\alpha$ be a complete set of local operators, including multi-traces.  Expand 
\begin{equation}
S(\delta) = \sum_\alpha \int d^d x \, \lambda^\alpha(x,\delta) {\cal A}_\alpha(x) \equiv S(\delta,\lambda)\,,
\end{equation}
where we allow for position-dependent couplings so as to have a generating functional for correlators.  Then 
the RG equation~(\ref{wilrg}) for the running is of the general form
\begin{equation}
\delta \partial_ \delta \lambda^\alpha(x, \delta) = -\beta^\alpha(x, \delta,\lambda) \,,
\end{equation}
and the statement of invariance with respect to cutoff is 
\begin{eqnarray}
0 = \frac{dZ}{d \delta}  = \delta \frac{\partial}{\partial \delta} \left\langle e^{-S(\delta,\lambda)} \right\rangle_ \delta - 
\int d^d x \, \beta^\alpha(x, \delta,\lambda) \frac{\delta}{\delta \lambda^\alpha(x)} \left\langle e^{-S(\lambda)} \right\rangle_ \delta \,,
\label{rg}
\end{eqnarray}
where the partial derivative acts only on the cutoff.
Further, using functional derivatives to introduce the correlator
\begin{equation}
{\cal C}_{\vec \alpha}(\vec y) \equiv {\cal C}_{\alpha_1,\ldots,\alpha_n}(y_1,\ldots,y_n) = \prod_{i=1}^n \frac{\delta}{\delta \lambda^{\alpha_i}(y_i)} \left\langle e^{-S(\delta,\lambda)} \right\rangle_ \delta \,,
\end{equation}
we have
\begin{equation}
0 = \delta \frac{\partial {\cal C}_{\vec \alpha}(\vec y)}{\partial \delta}  +
\int \!d^d x  \biggl( -\beta^\sigma(x, \delta,\lambda) \frac{\delta {\cal C}_{\vec \alpha}(\vec y)}{\delta \lambda^\sigma(x)}  + \sum_{i=1}^n 
 \Gamma_{\alpha_i}\!^\sigma(x,y_i, \delta,\lambda)
{\cal C}_{\alpha_1,\ldots,\sigma,\ldots,\alpha_n}
(y_1,\ldots,x,\ldots,y_n) \biggr) \,,
\end{equation}
where $\Gamma_{\alpha}\!^\sigma(x,y, \delta,\lambda) = \delta \beta^\sigma(x, \delta,\lambda) /\delta \lambda_\alpha(y)$.

In this discussion, we have taken a Euclidean metric, as is usual in studies of the renormalization group, but there appears to be no obstacle to having a Lorentzian metric on the Poincar\'e slices.

\subsection{Dynamical background}

Now we consider the general case, with dynamical metric. Because the metric is fixed on the AdS boundary one can in a neighborhood of the boundary fix the coordinates fully such that 
\begin{equation} \label{synchronous}
ds^2 = L^2 \frac{dz^2}{z^2} + h_{\mu\nu}(z,x) dx^\mu dx^\nu \,.
\end{equation}
These Fefferman-Graham coordinates have been employed extensively in the holographic renormalization program of Ref.~\cite{Henningson:1998gx}.  Once the coordinates are fixed, 
one can treat the $d$-dimensional metric $h_{\mu\nu}$ on the same footing as the scalar fields.\footnote{To be precise, we must add the Gibbons-Hawking surface term~\cite{Gibbons:1976ue}
 with opposite signs to the IR and UV parts of the bulk path integral,
 in order that these be well-defined.}  For pure Einstein gravity we get the Schr\"odinger equation
\begin{equation}
 \pm \kappa^2 \ell \partial_{\ell}\Psi(\ell,\tilde h_{\mu\nu}) = \int d^d
 x\left(-\frac{1}{\sqrt{\tilde h}}G_{\mu\nu\sigma\rho}(\tilde h) \tilde\pi_{\mu\nu} \tilde\pi_{\sigma\rho} - \sqrt{\tilde h} \tilde R^{(d)}\right)\Psi(\ell, \tilde h_{\mu\nu})\,,
\end{equation}
with the upper sign for the UV and the lower sign for the IR.  Tildes again denote the field restricted to $z = \ell$.  Here $\tilde\pi_{\mu\nu} = -i\kappa^2 \delta/\tilde h_{\mu\nu}$, 
$R^{(d)}(\tilde h)$ is the intrinsic curvature of a fixed $z$ slice, and 
\begin{eqnarray}
 G_{\mu\nu\sigma\rho} &\equiv& \frac{1}{2}\left(\tilde h_{\mu\sigma} \tilde h_{\nu\rho}+ \tilde h_{\nu\sigma} \tilde h_{\mu\rho} -\frac{2}{d-1} \tilde h_{\mu\nu} \tilde h_{\sigma\rho}\right).
\end{eqnarray}

This approach may raise concerns on two points, stability and coordinate invariance.  The Fefferman-Graham coordinates will in general break down at finite distance due to formation of caustics.\footnote{We thank A. Ashtekar for emphasizing this point.}  This is actually exacerbated by the AdS geometry: if we consider the radial geodesic $x^\mu = 0$, nearby geodesics $x^\mu = a^\mu z^2 + O(z^4)$ diverge quadratically.  We view this as an inessential complication.  One can think of this coordinate system as defining the surface of given $z_0$ as the set of points of distance $L \ln z_0/\epsilon$ from the surface $z = \epsilon$, where $\epsilon$ is in the asymptotic region~(\ref{adsmet}).  The coordinates $x^\mu$ are then assigned on the $z_0$ surface according to the closest point on the $\epsilon$ surface.  When there are caustics, the $z_0$ surface and the $x^\mu$ mapping are not smooth.  It should be possible to smooth these, for example by defining the distance from the $\epsilon$ surface in terms of some average distance to a disk of radius $z$.  We leave the details for future work, hopefully by others with better geometric tools.

More substantively, by fixing completely the lapse and shift we have lost the corresponding equations of motion, so that the state $\Psi_{\rm UV}$ does not satisfy the constraints of canonical gravity.\footnote{This point has also been made in the recent work~\cite{Lee:2009ij} on holographic and Wilsonian RG's.}  In our approach, we are foliating the bulk according to the lapse from the $\epsilon$ surface, and the integral over UV metrics is constrained to those of given total lapse and zero shift.  In the Wheeler-DeWitt (WdW) approach~\cite{DeWitt:1967yk}, one would integrate over all metrics with given inner and outer boundary, and identify a radial variable according to some intrinsic property of the geometry.  In this case the wavefunction would satisfy the constraints.

In most contexts it is the WdW wavefunction that is physical, but for the purpose of defining the Wilsonian action it is the coordinate-fixed wavefunction that is appropriate.  One sign of this is that the WdW wavefunction does not have the quasilocality expected of the Wilsonian action, because the action for the metric component $g_{zz}$ does not involve $z$-derivatives; we will work out explicitly the corresponding effect for the bulk gauge field in Sec.~3.2.  Another reason that this cannot work is that an ultralocal term in the metric would have to be of the form $\sqrt{ \tilde h}$ if the constraints are satisfied.  On the other hand, the Wilsonian action should contain arbitrarily ultralocal functions of $T_{\mu\nu}$, whose integral transform would be a correspondingly general function of $\tilde h_{\mu\nu}$.  Conceptually, the boundary QFT action is not coordinate invariant, and so neither should the cutoff action obtained by integrating out high energy fields.  Note that in most applications of gaussian normal coordinates there is no natural surface on which to begin, and so one must impose the constraints.  In AdS space the boundary provides such a surface.

For the IR path integral, there is no corresponding boundary surface and no condition to fix the coordinates globally.  One is integrating over all metrics with given outer boundary, and so $\Psi_{\rm IR}$ does satisfy the Hamiltonian and momentum constraints.  The equation of motion corresponding to the constraint, which is given by the matrix element of the constraint between $\Psi_{\rm IR}$ and $\Psi_{\rm UV}$, is therefore satisfied.

 As a result of the constraint, $\Psi_{\rm IR}$ has no direct dependence on $\ell$.  Rather,  for given $h_{\mu\nu}$ the IR path integral has an effective cutoff given by (\ref{cutoff}).
Insertions of the energy-momentum tensor are generated by derivatives with respect to $h_{\mu\nu}$, but we cannot linearize as in the scalar case~(\ref{ir}).  The RG equation is
\begin{equation}
0 = \frac{d}{d\ell}  \int {\cal D}\tilde h_{\mu\nu} {\cal D}\tilde\phi \,  \Psi_{\rm IR}(\tilde h_{\mu\nu},\tilde\phi)  \Psi_{\rm UV}(\ell,\tilde h_{\mu\nu},\tilde\phi) \,.
\end{equation}
When the metric is approximately classical, the UV amplitude will be peaked at a metric $h_{\mu\nu}(\ell)$.  This defines the cutoff~(\ref{cutoff}), and the integration over fluctuatations around $h_{\mu\nu}(\ell)$ generates multitrace interactions involving the energy-momentum tensor.

Incidentally, if one has a metric of finite total lapse in the IR direction, as with massive deformations of the QFT, most foliations will eventually break down.  Here there is a physical origin: when we pass through a massive threshold we must recast the RG in terms of new degrees of freedom appropriate to that scale.  More generally, the monotonicity of the energy with radius will break down, for example in multi-throated solutions corresponding to breaking of $U(N)$ to a product of smaller $U(M)$'s, and in semi-holographic situations where there are massless degrees of freedom in a region of finite warp factor~\cite{Faulkner:2010tq}.  Again, these represent thresholds in the RG.

\sect{Examples}

\subsection{Free scalars and double-trace flow}

It is interesting to work out in detail the case of a single free scalar, in the limit that backreaction is negligible.  We work in units $L=1$, and take a Poincar\'e invariant background, so $N = 1/z$, $h_{\mu\nu} = a^2(z)\delta_{\mu\nu}$, and
\begin{equation} \label{freescalarH}
H = \frac{1}{2} \int d^dx\, \ell^{-1} \Bigl\{ a^{-d} \tilde\pi^2 + a^{d} \bigl[ a^{-2}\delta^{\mu\nu} \partial_\mu \tilde\phi\partial_\nu \tilde\phi + m^2 \tilde\phi^2 - 2  d(d-1)\bigr] \Bigr\} \,.
\end{equation}
We further assume that the amplitudes are gaussian,
\begin{eqnarray}
\Psi_{\rm UV,IR}(\ell,\tilde\phi) &=& e^{- \kappa^{-2} W_{\rm UV,IR}(\ell,\tilde\phi)} \,, \nonumber\\
W(\ell,\tilde\phi) &=& C(\ell) + \frac{1}{2} \int\frac{d^dk}{(2\pi)^d} F(k,\ell) \left(\tilde\phi-\bar\phi(\ell)\right)_{-k}\left(\tilde\phi-\bar\phi(\ell)\right)_k  \,.   \label{wgau}
\end{eqnarray}
In the IR this follows dynamically, while in the UV it is an assumption about the initial conditions on the flow, which could be relaxed.  We have allowed for a possible translation-breaking source for the linear term, but not for the quadratic one: the subscript $k$ is for Fourier components of fields and sources, while the argument $k$ is for kernels.

Then
\begin{align}\label{freescalareqns}
 \pm \ell \partial_\ell F &= - a^{-d} F^2 + {a^{d}} ( a^{-2} k^2 + m^2) \,,
\\
\pm \ell \partial_\ell \bar\phi_{k} &=  -( a^{-2} k^2 + m^2) \bar\phi_{k}/ {a^{-d}}F \,, \label{fphi}
\end{align}
with the upper sign for the UV and the lower sign for the IR.  For the normalization constant,
\begin{equation}
\pm \ell\partial_\ell C = \int d^d x \, a^{d} d(d-1) + \frac{1}{2} \int\frac{d^dk}{(2\pi)^d} 
\left[
a^{-d} F^2 \kappa^2 + a^{d} ( a^{-2} k^2 + m^2) \bar\phi_{k}  \bar\phi_{-k}
 \right] \,.
\end{equation}
For this quadratic system the full Schr\"odinger equation differs from the HJ equation only by the $O(\kappa^2)$ term in $C$.

The center of the gaussian, $\bar\phi_k(\ell)$, is not a solution to the bulk field equation, but it and  $F(\ell,k)$ can be written
\begin{equation} \label{Fsoln}
 F = \pm \frac{a^d \ell \partial_\ell \vartheta}{\vartheta}\,, \quad \bar\phi_k = \frac{B_k}{a^d \ell\partial_\ell \vartheta} \,, 
\end{equation}
where $\vartheta(\ell,k)$ does solve the massive Klein-Gordon equation 
\begin{equation}
 \ell\partial_\ell (a^d \ell\partial_\ell \vartheta) = a^d(a^{-2}k^2+m^2) \vartheta\,.
\end{equation}
We now specialize to the case of pure AdS, $a=\ell^{-1}$, for which $\vartheta $ is a Bessel function of order $\nu = \sqrt{d^2/4+m^2}$.  It is convenient to write it as
\begin{equation}
 \vartheta(k,\ell)= Y_+(k,\ell) + A(k) Y_-(k,\ell) \,,
 \end{equation}
 in terms of$\,$\footnote{For integer $\nu$ we will need a different prescription, below.}
\begin{equation}
Y_{\pm}(k,z) = z^{d/2} I_{\pm \nu}(kz) \Gamma(1 \pm \nu) (k/2)^{\mp \nu}\,.
\end{equation}
The set of flows is thus parameterized by two functions $A(k)$, $B_k$.

\subsubsection{UV amplitude}

First let us cast the evolution equations in the form of an RG flow.
According to Eq.~(\ref{gauss}), $\Psi_{\rm UV}$ corresponds to the action
\begin{equation}
s_{\ell} = \int \frac{d^dk}{(2\pi)^d} \left( -\bar\phi_{{\rm UV,}-k}(\ell) {\cal O}_k - \frac{1}{2F_{\rm UV}(k,\ell)}
{\cal O}_k {\cal O}_{-k} \right) \,.
\end{equation}
Define single- and double-trace couplings
\begin{equation}
g_{k}(\ell) =  -\bar\phi_{{\rm UV,}-k}(\ell) \,,\quad h(k,\ell) = - \frac{1}{F_{\rm UV}(k,\ell) \ell^{d}} \,.
\end{equation}
The evolution equations~(\ref{fphi}) become 
\begin{eqnarray}
\ell \partial_\ell g_{k} &=&   ( \ell^{2} k^2 + m^2) h(k) g_{k} \equiv - \beta_{g_k}
\,,\nonumber\\
\ell \partial_\ell h(k) &=&  - 1 - d h(k) + ( \ell^{2} k^2 + m^2) h^2(k) \equiv - \beta_{h(k)}\,.  \label{beta}
\end{eqnarray} 
We can also interpret the evolution of $C$ as a running of the unit operator,
\begin{equation}
\beta_1 =  -\int d^d x \, \ell^{-d} d(d-1) - \frac{1}{2} \int\frac{d^dk}{(2\pi)^d} 
 \ell^{-d} ( \ell^{2} k^2 + m^2) g_{k}  g_{-k} + O(\kappa^2) \,.
\end{equation}

At momenta long compared to the cutoff, where the $\ell^2 k^2$ term can be neglected, the double-trace coupling has two fixed points,
\begin{equation}
h(k) = \frac{\Delta_\pm}{m^2} = - \frac{1}{\Delta_\mp} \,,\quad \Delta_\pm = \frac{d}{2} \pm \nu \,.
\end{equation}
At these fixed points,
\begin{equation}
g_k \sim \ell^{\Delta_\pm} \,.
\end{equation}
Correspondingly, the single trace operator $\cal O$ has dimension $d-\Delta_\pm = \Delta_\mp$: the lower sign is the standard quantization, and the upper sign is the alternate quantization~\cite{Klebanov:1999tb}.
 
Also when $k \ell \ll 1$, we can take the leading behaviors of the Bessel functions, 
\begin{equation}
\vartheta_{\rm UV}(k,z) \approx  z^{\nu + d/2} + A_{\rm UV}(k) z^{-\nu + d/2} \,,
 \end{equation}
so that
\begin{equation}
h(k) = -\frac{A_{\rm UV}(k) + \ell^{2\nu}  }{ \Delta_- A_{\rm UV}(k) + \Delta_+ \ell^{2\nu}  } \,.
\end{equation}
For generic $A_{\rm UV}(k)$ this describes the flow~\cite{Witten:2001ua,Berkooz:2002ug} from the alternate quantization in the UV to the standard quantization in the IR.  The standard conformal quantization is $A_{\rm UV}(k) = 0$ and the alternate conformal quantization is $A_{\rm UV}(k) = \infty$.  Recall that for $\nu \geq 1$ only the standard quantization gives a unitary theory all the way into the UV, so we focus on $\nu < 1$; we could also consider the flow at general $\nu$ in an effective theory with a maximum UV scale.

The UV boundary conditions on the fields are generally parameterized as
\begin{equation}
\phi_{k} \sim \alpha_k z^{\Delta_-} + \beta_k z^{\Delta_+} \,.
\label{asymp}
\end{equation}
Our restriction to gaussian $\Psi_{\rm UV}$ implies a linear relation
\begin{equation}
\alpha_k = f(k) \beta_k + j_k \,. \label{linab}
\end{equation}
In order to relate this parameterization to the general flow above, we can directly evaluate the path integral, which is extremized by $\phi^{\rm cl}_{{\rm UV},k}(\ell)$, which has the asymptotic behavior~(\ref{asymp}) at $z=0$ and is equal to $\tilde\phi$ at $z = \ell$. The solution to the classical equation of motion
 \begin{equation}
 z\partial_z (a^d z\partial_z \phi) = a^d(a^{-2}k^2+m^2) \phi
 \label{bulkKG}
 \end{equation}
 with given boundary conditions is
\begin{equation}
\phi^{\rm cl}_{{\rm UV},k}(z) = \tilde\phi_k \frac{Y_+(k,z) + f(k) Y_-(k,z)}{Y_+(k,\ell) + f(k) Y_-(k,\ell)}
+ j_k  \frac{Y_+(k,\ell) Y_-(k,z) - Y_-(k,\ell) Y_+(k,z)}{Y_+(k,\ell) + f(k) Y_-(k,\ell)} \,. \label{phicl}
\end{equation}
Using the classical equation of motion and integrating by parts gives
\begin{equation}
W_{\rm UV}(\ell,\tilde\phi) = \frac{1}{2} \int\frac{d^dk}{(2\pi)^d}\left[ \ell^{1-d}\phi^{\rm cl}_{{\rm UV},k}(\ell) \partial_\ell \phi^{\rm cl}_{{\rm UV},-k}(\ell) - 2\nu j_k \beta_{-k} \right]
 \,. \label{wuv}
\end{equation}
We have included a surface term, $\frac{1}{2} \epsilon^{1-d} \phi_k(\epsilon) \partial_\epsilon \phi_{-k}(\epsilon) - \frac{1}{2} \epsilon^{-\Delta_+} j_k (\epsilon \partial_\epsilon - \Delta_-) \phi_{-k}(\epsilon)$, as needed to give a good variational principle with the boundary condition~(\ref{linab}).
Inserting the classical solution, we can reduce this to the earlier form~(\ref{Fsoln}), with
\begin{equation}
A(k) = f(k) \,,\quad
B_k = -2\nu j_k  \,.
\end{equation}

The $Y_{\pm}(k,\ell)$ are by construction entire functions of $k$.\footnote{In order to retain this property for integer $\nu$, i.e. $\nu \to n$, we need in this case to replace $Y_{-}(k,z)$ with $\lim_{\nu \to n}
z^{d/2} (I_{- \nu}(kz) - k^{2(n-\nu)} I_{\nu}(kz)) \Gamma(- \nu) (k/2)^{ \nu}$.}
The UV amplitude depends only on $Y_{\pm}(k,\ell)$ and its derivative, and on the initial data $j_k, f(k)$.  Therefore if the latter are are localized (analytic around $k=0$) then so is the WIlsonian action, as expected. 
We can then expand
\begin{equation}
h(k) = \sum_{n=0}^\infty h_n k^{2n}\,,
\end{equation}
then $h_n$ is the coefficient of $\frac{1}{2} {\cal O}(x) (-\partial^2)^n  {\cal O}(x)$ in the Lagrangian.  The RG equation becomes
\begin{equation}
\ell \partial_\ell h_n = - \delta_{n0} - d h_n + m^2 \sum_{m=0}^n h_m h_{n-m}
+ \ell^2 \sum_{m=0}^{n-1} h_m h_{n-m-1} \,.
\label{momex}
\end{equation}
Note that the equation for $h_n$ involves  on the right only $h_{n'}$ with $n' \leq n$, allowing an iterative solution.  This iterative structure in the momentum expansion is a property of the planar limit in the interacting case as well.

\subsubsection{IR amplitude}
 
The infrared amplitude solves the evolution equations~(\ref{fphi}) with the lower sign.   In the path integral, the condition that the classical solution not blow up as $z \to \infty$ fixes 
\begin{equation}
\phi^{\rm cl}_{{\rm IR},k}(z) = \tilde\phi_k  z^{d/2} \frac{K_\nu(kz)}{K_\nu(k\ell)}\,, \quad F_{\rm IR} = - \frac{ \ell \partial_\ell (\ell^{d/2} K_\nu(k\ell))}{\ell^{3d/2} K_\nu(k\ell)} \,,
\label{fir}
\end{equation}
so $\vartheta_{\rm IR}(k,z) = z^{d/2}K_\nu(kz)$ and $\bar\phi_{{\rm IR},k} = 0$.
The amplitude is
\begin{equation} \label{freescalarIRamplitude}
\Psi_{\rm IR} = e^{-W_{\rm IR}} \,,\quad W_{\rm IR}(\ell,\tilde\phi) = \frac{1}{2} \int\frac{d^dk}{(2\pi)^d} F_{\rm IR}(k,\ell) \tilde\phi_{-k} \tilde\phi_k \,.
\end{equation}

To see how the RG acts on this, we have from~(\ref{ir})
\begin{equation}
\Psi_{\rm IR}(\ell,\tilde\phi) =  \left\langle e^{-S(-\tilde\phi,0)} \right\rangle_\ell \,,
\end{equation}
in terms of the action $S(g,h)$ with single- and double-trace couplings.  At $h=0$ we simply have $\beta_g = 0$ and $\beta_h = 1$.
The RG~(\ref{rg}) then takes the form
\begin{equation}
\label{wilir}
\ell \frac{\partial}{\partial\ell} \Psi_{\rm IR}(\ell,\tilde\phi) = \beta_1\Psi_{\rm IR}(\ell,\tilde\phi) +  \int\frac{d^dk}{(2\pi)^d} \frac{\delta}{\delta h(k)}  \left\langle e^{-S(-\tilde\phi,h)} \right\rangle_\ell\Bigr|_{h=0} \,.
\end{equation}
The integrand of the second term on the right can be rewritten
\begin{equation}
\frac{1}{2} \left\langle {\cal O}_k {\cal O}_{-k} e^{-S(-\tilde\phi,0)} \right\rangle_\ell
= \frac{1}{2} \frac{\delta}{\delta\tilde\phi_k}\frac{\delta}{\delta\tilde\phi_{-k}} \Psi_{\rm IR}(\ell,\tilde\phi) \,,
\end{equation}
and so we recover the Schr\"odinger equation.  In the HJ approximation, the RHS becomes
\begin{equation} \label{HJIR}
\frac{1}{2} \frac{\delta W_{\rm IR}}{\delta\tilde\phi_k}\frac{\delta  W_{\rm IR}}{\delta\tilde\phi_{-k}} e^{- W_{\rm IR}} \,.
\end{equation}

The IR amplitude is generated by integrating out long wavelengths, and so is not localized.  Using 
$K_\nu(kz)= {\Gamma(\nu)\Gamma(1-\nu)}(I_{-\nu}(kz)-I_\nu(kz))/2$, we have
\begin{equation} \label{IRsoln}
 F_{\rm IR}(k,\ell) =-\frac{1}{\ell^d}\left(\frac{d}{2}-\nu + k\ell\frac{I_{1-\nu}(k\ell)-I_{1+\nu}(k\ell)}{I_{-\nu}(k\ell)-I_{\nu}(k\ell)} \right) \,.
\end{equation}
In the asymptotic expansion at small $k$, one obtains terms of the form $k^{2m + 2n\nu}$.
The local part, defined by the terms of the form $k^{2m}$, is
\begin{equation}
 F_{\rm IR, loc}(k,\ell) =-\frac{1}{\ell^d}\left(\frac{d}{2}-\nu + k\ell\frac{I_{1-\nu}(k\ell)}{I_{-\nu}(k\ell)}\right) \,.
\end{equation}
This has no particular relation to the UV amplitude, which is also local, since the latter depends on the UV boundary conditions.  In identifying the local part, we have ignored additional integer powers that appear when $\nu$ is rational; for more general flows, picking out an part analytic in $k$ may not be well defined.  However, the ultralocal $k^0$ term is unambiguous if $\nu > 0$, and in the present case is just
\begin{equation} \label{freescalarultralocalIR}
 F_{\rm IR, loc}(0,\ell) =-\frac{\Delta_-}{\ell^d}\,.
\end{equation}
The leading nonlocal behavior at small $k$ is
\begin{equation} \label{freescalarnlIR}
 F_{\rm IR, nl}(0,\ell) \approx {\rm const} \times k^{2\nu} \ell^{-2\Delta_-}
 \end{equation}

\subsection{Bulk gauge field}

Our next example is a bulk $U(1)$ gauge field $A^\mu$ in a fixed Poincar\'e invariant background geometry, which must couple to a conserved current $j^\mu(M(x))$ in the boundary theory.
Integrating by parts along the $x$ directions the action can be written 
\begin{equation}
\mathcal{S} = -\frac{1}{2} \int d^d x\, dz \,\frac{a^{d}}{z} \left\{ \frac{z^2}{a^2}\delta^{\mu\nu} (\partial_z A_\mu - \partial_\mu A_z)(\partial_z A_\nu - \partial_\nu A_z) - \frac{1}{a^{4}} A_{\mu} P^{\mu\nu} \partial^2 A_{\nu} \right\}\,,
\end{equation}
where $P^{\mu\nu} = \delta^{\mu\nu} - \partial^\mu \partial^\nu/\partial^2$ is a projection operator.
We will first deal with the gauge symmetry by fixing radial gauge, $A_z = 0$.  Normally this leaves a residual invariance under $z$-independent gauge transformations, but in AdS space the gauge parameter is required to vanish at the boundary (in the CFT the $U(1)$ is a global symmetry), so the gauge symmetry is fully fixed.

We can then write a Hamiltonian directly
\begin{equation}
H_{\rm rad} = \frac{1}{2z}\int d^d x \left( {a^{2-d}} \tilde\pi^2 -a^{d-4} \tilde A_\mu P^{\mu\nu} \partial^2 \tilde A_\nu \right) \,,
\end{equation}
giving the (momentum-space) Schrodinger equation
\begin{equation} \label{bulkgaugeSchrodinger}
 \mp \kappa^2\ell \partial_{\ell} \Psi(\ell, \tilde A_\mu) = \frac{1}{2}\left(- \frac{\kappa^4 \eta_{\mu\nu} }{a^{d-2}} \frac{\delta^2}{\delta \tilde A_\mu \delta \tilde A_\nu} +a^{d-4}k^2  P^{\mu\nu}(k) \tilde A_\mu \tilde A_\nu \right)\Psi(\ell, \tilde A_\mu) \,,
\end{equation}
where the upper/lower sign is for the \mbox{UV/IR}. The Schr\"odinger equation separates in  the longitudinal and transverse variables $\tilde A^{\rm T}_\mu = P_{\mu}^\nu \tilde A^{\vphantom T}_\nu$, $\tilde A^{\rm L}_\nu = \tilde A^{\vphantom T}_\nu - \tilde A^{\rm T}_\nu$,
\begin{equation}
\Psi(\ell, \tilde A^{\vphantom T}_\mu) = \Psi^{\rm L}(\ell, \tilde A^{\rm L}_\mu) \Psi^{\rm T}(\ell, \tilde A^{\rm T}_\mu) \equiv e^{-\kappa^{-2}(W^{\rm L}+W^{\rm T})} \,.
\end{equation}

We first obtain $\Psi_{\rm UV}$ by direct evaluation of the path integral with $A_\mu(\ell,x) = \tilde A_\mu(x)$ and $A_\mu(0,x) \equiv A_{\rm b,\mu}(x) $ fixed.  The latter is the same as the classical gauge field coupled to the \mbox{CFT.} The equations of motion in radial gauge reduce to
\begin{equation} \label{TLEOM}
z\partial_z (a^{d-2}z\partial_z A^{\rm T}_\mu) = a^{d-4}\partial^2 A^{\rm T}_\nu\,, \qquad \partial_z ( a^{d-2}z\partial_z A^{\rm L}_\mu) = 0 \,.
\end{equation}
The transverse part obeys the scalar Klein-Gordon equation~(\ref{bulkKG}) with the replacements $m=0$ and $d\to d-2$, while the longitudinal part is given by
\begin{equation}
A^{\rm L}_\mu(z,x) = \left(\tilde A^{\rm L}_\mu(x) - A^{\rm L}_{\rm b,\mu}(x) \right) \frac{\sigma(z)}{\sigma(\ell)} + A^{\rm L}_{\rm b,\mu}(x)\,, \quad \sigma(z) \equiv \int_\epsilon^z \frac{dz'}{z'} a^{2-d}(z') \,.
\end{equation}
The classical action is then
\begin{equation}
\mathcal{S}^{\rm cl} =  \lim_{\rm \epsilon \to 0} -\frac{1}{2}\int d^dx \, a^{d-2}z A^{\rm cl,\mu}\partial_z A^{\rm cl}_{\mu} \bigg|^\ell_\epsilon 
= W^{\rm L} + W^{\rm T}
\end{equation}
The longitudinal part is
\begin{equation}
W_{\rm UV}^{\rm L} = -\frac{1}{2}\frac{(\tilde A_\mu^{\rm L}(x) - A_{\rm b,\mu}^{\rm L}(x))^2}{\sigma(\ell)} \,.
\end{equation}
The transverse part is
\begin{align}
W_{\rm UV}^{\rm T} &= C(\ell) + \frac{1}{2}\int \frac{d^d k}{(2\pi)^d} F(k,\ell) P^{\mu\nu}(k) (\tilde A^{\rm T}_\mu - \bar A^{\rm T}_\mu(\ell))_{-k} (\tilde A^{\rm T}_\nu -\bar A^{\rm T}_\nu(\ell))_k \,,
\end{align}
where 
\begin{equation}
F = \frac{a^{d-2} \ell\partial_\ell \vartheta_+}{\vartheta_+}, \qquad \bar A^{\rm T}_\mu = \frac{A^{\rm T}_{{\rm b},\mu k}}{a^{d-2}\ell \partial_\ell \vartheta_+} \,.
\end{equation}
Here $\vartheta_+(z)$ is the solution to the massive Klein-Gordon equation at $m = 0$ and $d \to d-2$ such that $\lim_{z\to 0} a^{d-2}\ell \partial_\ell \vartheta_+ = 1$.

It is clear from the path integral construction that the total classical action is localized in the boundary values of the fields.   This is not manifest in the final form, because the projection $P_\mu^\nu$ makes the transverse and longitudinal pieces separately nonlocal: each piece separately is $1/k^2$ times an analytic function of $k^2$, and a generic solution to the Schr\"odinger equation would be nonlocal.   One can verify that the pole cancels in the sum by noting that at $k = 0$, $\vartheta_+(z) = \sigma(z)$ and $a^{d-2}z \partial_z \vartheta_+ $ is identically 1, so that the nonderivative part of the action is simply
\begin{equation}
W_{\rm UV}^{\rm L} + W_{\rm UV}^{\rm T} = -\frac{1}{2}\frac{(\tilde A_\mu(x) - A_{\rm b,\mu}(x))^2}{\sigma(\ell)} \,.
\end{equation}
The integral transform, which is the Wilsonian action, therefore contains a a $j_\mu j^\mu$ term and an $A_{\rm b,\mu} j^\mu$ term.

If we had fixed $A_z$ to some generic functional form, then by a gauge transformation we can see that the action would be the same as above with the replacement
\begin{equation}
A^{\rm L}_\mu(\ell,x) \to A^{\rm L}_\mu(\ell,x) - \partial_\mu\Theta(x)\,, \quad
\Theta(x) = \int_0^\ell dz\,A_z(z,x) \,.
\end{equation} 
If we integrated freely over $A_z$, the result would be a gauge volume times an integral over $\Theta(x)$, with a result independent of $A^{\rm L}$.  That is, the UV amplitude would now satisfy the constraint
\begin{equation}
 \partial_\mu \frac{\delta}{\delta \tilde A_\mu} \Psi_{\rm UV} = 0 \,.
\end{equation}
In canonical quantization, one would normally be interested only in such physical states, but for our purposes this cannot give rise to a Wilsonian action: eliminating the dependence on $A^{\rm L}$ necessarily leaves the amplitude nonlocal.  Effectively, the Wilson action must remember that the $U(1)$ symmetry was a global symmetry in the UV, not a gauge symmetry.

All of these observations are parallel to our discussion of the coordinate choice in Sec.~2.2: the radial gauge has the same properties as the gaussian normal gauge, and gives rise to a wavefunction that does not satisfy the constraints, but gives a Wilson action with the expected properties.  The $A_z$-integrated amplitude is analogous to the Wheeler-DeWitt wavefunction and does not lead to a local action.\footnote{It is interesting to note the recent paper~\cite{Nickel:2010pr}, in which these nonlocal effects are precisely the interesting part of the physics.  In our way of organizing things these are part of the IR amplitude, coming from the relative gauge or coordinate transformation between the horizon and the cutoff radius.}

The longitudinal part of the IR amplitude is 
\begin{equation}
W_{\rm IR}^{\rm L} = -\frac{1}{2}\frac{(\tilde A_\mu^{\rm L}(x) - A_{\mu}^{\rm L}(z_0,x))^2}{\sigma_{\rm IR}(\ell)} \,, \quad \sigma_{\rm IR}(z) \equiv \int_z^{z_0} \frac{dz'}{z'} a^{2-d}(z')  \,.
\end{equation}
When there is a horizon ($z_0 =  \infty$) then $\sigma_{\rm IR}(\ell)= \infty$ and the longitudinal part vanishes (except for $d = 1$, but this is a rather special case).  When there is a mass gap (so $z_0$ is finite),  $\sigma_{\rm IR}(\ell)$ is finite but we must integrate over the gauge field on the lower boundary, and the longitudinal amplitude is again trivial.  Thus the IR amplitude always satisfies the constraint.

\subsection{Domain-wall flow}

Now we consider an interacting scalar theory, taking for simplicity a single field.  In this subsection we assume that the backreaction on the metric is small, in the next we discuss its inclusion.  We focus on the momentum-independent terms in the action.  These satisfy closed equations, as we have already seen in the example~(\ref{momex}).  Thus, writing
\begin{equation}
W_{\rm UV} = \int d^dx\, \ell^d w(\ell,\tilde\phi) \,,
\end{equation}
we have the HJ equation
\begin{equation}
\ell \partial_\ell w(\ell,\tilde\phi) = - d w(\ell,\tilde\phi) - \frac{1}{2} w'(\ell,\tilde\phi) w'(\ell,\tilde\phi) + V(\tilde\phi) \,.
\end{equation}
Suppose that $V$ has a stationary point $\phi_0$,
\begin{equation}
V(\tilde\phi) = v_0 + \sum_{n=2}^\infty v_n (\tilde\phi- \phi_0)^n \,.
\end{equation}
Then there are scale invariant solutions
\begin{equation}
w(\tilde\phi) = w_0 + \sum_{n=2}^\infty w_n (\tilde\phi- \phi_0)^n \,,
\end{equation}
where
\begin{eqnarray}
d w_0 &=& v_0  \,, \nonumber\\
dw_2 &=& -2 w_2^2 + v_2 \,, \nonumber\\
(d + w_2) w_m &=& -\frac{1}{2} \sum_{k=3}^{m} k(m-k+2) w_k w_{m-k+2} + v_m \,,\quad m\geq 3 \,.
\label{iter}
\end{eqnarray}
Note that the equation for $w_2$ is independent of the other $w_m$, and quadratic.  This just produces the standard and alternate quantizations, subject to the usual restrictions.  The remaining equations then determine the higher $w_m$ iteratively.   

Flows that approach one of the extrema of $V$ in the UV correspond to perturbations by relevant operators.  Much of the literature on the holographic renormalization group focuses on flows that interpolate between different extrema of $V$.  In our framework, we look for a solution of the form
\begin{equation}
w(\tilde\phi,\ell) = \sum_{n=0}^\infty w_n(\ell) (\tilde\phi- \phi_{0}(\ell))^n \,.
\end{equation}
This form is redundant, because $\phi_{0}(\ell)$ can be set to an arbitrary function by redefinition of the $w_n$'s, but it is useful to present it in this way.
Taylor expanding the HJ equation around $\phi_0$ gives
\begin{equation}
(\ell \partial_\ell + d) w_m = (m+1) w_{m+1} \ell \partial_\ell\phi_0
-\frac{1}{2} \sum_{k=1}^{m+1} k(m-k+2) w_k w_{m-k+2} + \frac{1}{m!} \partial^mV(\phi_0).
\end{equation}
The first three equations are
\begin{align}
\label{w1}
(\ell \partial_\ell + d) w_1 &= 2 w_{2} (\ell \partial_\ell\phi_0- w_1) +  \partial V(\phi_0),\\
\label{w2}
(\ell \partial_\ell + d) w_2 &= 3 w_3(\ell \partial_\ell\phi_0- w_1) - 2 w_2^2 + \frac{1}{2} \partial^2 V(\phi_0),\\
\label{w3}
(\ell \partial_\ell + d) w_3 &= 4 w_4(\ell \partial_\ell\phi_0- w_1) - 6 w_2 w_3 + \frac{1}{6} \partial^3 V(\phi_0).
\end{align}

We can obtain the same iterative structure~(\ref{iter}) as in the conformal case if we fix the redundancy noted above by requiring
\begin{equation}
w_1 = \ell \partial_\ell \phi_{\rm 0} \, .
\end{equation}
The $w_1$ equation is then independent of $w_2$, and requires that $\phi_0(\ell)$ satisfies the classical field equation,
\begin{equation}
  (\ell \partial_\ell + d) \ell \partial_\ell \phi_{0}= \partial V(\phi_{0}) \,.
\end{equation}
In the equations for the higher $w_m$, the next term $w_{m+1}$ enters multiplied by $w_1 - \Lambda \partial_\Lambda \phi_{0} = 0$, so each successive equation closes.  

For the case of a domain wall solution~\cite{Klebanov:1998hh}, where $\phi_0$ approaches distinct extrema of $V$ at $\ell \to \pm\infty$, this asymptotes to the conformal solution at each end.
The quadratic term $w_2$, adjusts in response to the flow of $w_1$ (through $\phi_0$), and this is necessary to obtain the evolution of the connected two point-functions, as calculated directly in Ref.~\cite{Porrati:1999ew}.

In particular, Eq.~\eqref{w2} is now solved by $w_2 = \ell\partial_\ell \psi_{2}/2\psi_{2}$, where
$(\ell \partial_\ell + d) \ell \partial_\ell \psi_{2}  = \psi_{2} \partial^2 V(\phi_0)$.
For a quadratic potential, discussed in the previous section, $\psi_{2}=\phi_0$. If we then shift $\phi \to \phi + w_1/2w_2$ to eliminate the linear term we find that the quadratic term is centered at zero, i.e. $\bar\phi=0$.

In the flow studied in Sec.~3.1 we chose instead $w_1 = 0$.  The $w_1$ equation then becomes
\begin{equation}
 \ell\partial_\ell \phi_0 = -\partial V / 2 w_2 \,.
\end{equation}
Here $\phi_0$ is not a solution to the field equation but obeys the same equation~(\ref{fphi}) as $\bar\phi$.   However, the equation for $w_2$ depends on $w_3$, which cannot be zero for a nonquadratic potential, and so does not close as in Eq.~(\ref{freescalareqns}).
In this form, the $w_i$ do not decouple from the $w_{i+1}$.

For the Wilsonian action, again consider the momentum-independent terms
\begin{eqnarray}
s(\ell) &=& \int d^dx\,\sigma(\ell, \mathcal{O}) \nonumber\\
\partial_\ell \sigma &=& -\frac{1}{2} \mathcal{O}^2 + V(-\sigma') \,.
\end{eqnarray}
Expanding
\begin{equation}
V(\tilde\phi) = \sum_{n=0}^\infty v_n \tilde \phi^n \,, \quad \sigma(\mathcal{O}) =  \sum_{n=0}^\infty \sigma_n \mathcal{O}^n \,,
\end{equation}
one finds that 
\begin{equation}
 \ell\partial_\ell \sigma_n = (n+1) V'(\sigma_1) \sigma_{n+1} + \ldots\,, 
\end{equation}
where the ellipsis represents terms that contain only $\sigma_{n'}$ such that $n' \neq n$.  Unlike the 
equations for $W$, these are not in general iterative in $n$.

\subsection{Backreaction}

Now let us consider the backreaction on the metric.  To emphasize the main points we consider the case without scalars, with negative cosmological constant $V = -d(d-1)/L^2$.  For
\begin{equation}
W_{\rm UV} = \int d^dx\, \ell^d w(\ell,a) \,,
\end{equation}
with $a^{2d} = \det h$, the HJ equation is
\begin{equation}
\ell \partial_\ell w(\ell,a) = \frac{a^{2-d}}{4d(d-1)} (\partial_a w)^2 - d (d-1) a^d  \,.
\end{equation}
The solution, given by the classical action with $a \to L/\ell$ as $\ell \to 0$, is
\begin{equation}
w(\ell,a) = 4(d-1) L^d /\ell^d - 2(d-1)(a^{d/2} - 2 L^{d/2}/\ell^{d/2})^2 \,.
\end{equation} 
Note the wrong-sign quadratic term, a manifestation of the sign problem of Euclidean gravity~\cite{Gibbons:1978ac}.  Roughly speaking, the $a$ contour must be rotated into the imaginary direction at the saddle point.  Again, the UV action does not satisfy the constraint, which would allow only a term $a^d$.

For a general domain wall solution with back-reaction, ($\phi_{0}(\ell), a_{0}(\ell)$), one can simply treat the scale factor on the same footing as the scalars, and so extend the previous discussion.

\sect{Relations between RG formalisms}

\subsection{Projections}

If we are applying this formalism to the full bulk theory including excited string states, we must correspondingly include in the field theory all operators, and their flow is described by the Wilsonian \mbox{RG}.   In comparing different forms of the RG, it is important to distinguish this full Wilsonian RG from various forms where the flow has been projected onto a subspace.

The most familiar projection is the RG in renormalizable theories.  In the Wilsonian framework, nonrenormalizable couplings correspond to directions where the flow of the action is strongly convergent.\footnote{Note that the evolution equations~(\ref{evolve}) require that we evolve the action toward the IR, and the cut-off amplitude $\Psi_{\rm IR}$ toward the UV, for typical Hamiltonians that are unbounded above.}  Projecting onto the remaining directions, one obtains the flow of the nearly-marginal operators, those for which $\Delta - d$ is parametrically small: this is the renormalizable flow, described by the Callan-Symanzik equation.  Superrenormalizable couplings correspond to directions in which the flow is diverging, and these must be suppressed by tuning or by symmetries to obtain a flow over a wide range of scales.  In the bulk, this projection corresponds to an effective theory of nearly massless fields, $m^2 L^2 \ll 1$.  

It is important to note that this greatly changes the form of the RG.  For example, in the framework of Ref.~\cite{Polchinski:1983gv}, the full RG contains only a tree level term which is quadratic in the couplings, and a one-loop term which is linear in the couplings, and these are explicitly known.  The renormalizable flow, on the other hand, contains terms of all orders in the coupling, which must be calculated order by order.  To see how this comes about, consider the model equations~\cite{Polchinski:1983gv}
\begin{equation}
\ell \partial_\ell \lambda_4 = \lambda_6 \,,\quad \ell \partial_\ell \lambda_6 = - 2\lambda_6 +\lambda_4^2 \,. \label{toyrg}
\end{equation}
Here $\lambda_4$ is a nearly marginal coupling, and $\lambda_6$ is irrelevant.  As we flow toward the IR the latter approaches a value that depends only on $\lambda_4$ and not on the initial conditions.  In this simple example we can immediately write the resulting equation for $\lambda_4$ by taking $\ell \partial_\ell$ of the first of Eqs.~(\ref{toyrg}) and then using the second,
\begin{equation}
2 \ell \partial_\ell \lambda_4 = \lambda_4^2 - (\ell \partial_\ell)^2 \lambda_4 \,.
\end{equation}
This is to be solved in a `slow-roll' approximation, where the scale derivatives are of order $\lambda_4$ and so the second term on the right is subleading.  Inserting the leading approximation $2 \ell \partial_\ell \lambda_4 = \lambda_4^2 + O(\lambda_4^3)$ into this term gives 
\begin{equation}
2 \ell \partial_\ell \lambda_4 = \lambda_4^2 - \frac{1}{2}\ell \partial_\ell(\lambda_4^2) + O(\lambda_4^4)
=   \lambda_4^2 - \frac{1}{2} \lambda_4^3 + O(\lambda_4^4)\,.
\end{equation}
Continuing the iteration gives the infinite series typical of renormalizable theories.

In theories having a gravity dual, there is another natural truncation.  In this case there is a set of operators whose dimensions are of order one, dual to states whose masses are of order the AdS scale, and there are operators whose dimensions are parametrically large, dual to string scale (or heavier) states.  These ultra-high dimension fields represent directions in which the flow is converging extremely rapidly, and one can project onto the relatively slower flow of operators of dimension $O(1)$.  In the bulk, this corresponds to truncating to the effective supergravity theory including the dependence of the fields on the compact directions (when these are at the AdS scale).  Indeed, our whole framework of separating the bulk path integral radially into two parts is much more natural in this effective theory than in the full string theory.  Thus it will be of interest to study this `supergravity' truncation of the Wilsonian RG~\cite{Heemskerk:2009pn}. 


\subsection{Holographic RG}

The correspondence between radius and energy of course goes back to the first work on AdS/CFT~\cite{Maldacena:1997re,Gubser:1998bc,Witten:1998qj}, and many papers have explored the
 analogy between a radial cutoff and the Wilson \mbox{RG}.  Some of our results have appeared in one form or another, but our main result~(\ref{wilrg}) does not seem to have been noted, and much of the holographic RG literature seems rather different in form, in particular dealing exclusively on the single-trace operators.

The widely-employed formalism of Ref.~\cite{de Boer:1999xf}, focuses on the low energy amplitude $\Psi_{\rm IR}$ and does not consider the Wilsonian action as we have identified it.  This amplitude satisfies the evolution equation
\begin{equation}
\partial_\ell \Psi_{\rm IR}(\ell,\tilde\phi) = H(\tilde\phi, \tilde\pi) \Psi_{\rm IR}(\ell,\tilde\phi) \,.
\label{irevol}
\end{equation}
It is a function only of the $\tilde\phi^i$, which are in one-to-one correspondence with the single-trace operators, and so these are naturally interpreted as couplings in a purely single-trace action.  However, Eq.~(\ref{irevol}) does not have the form of an RG: the latter involves only first derivatives with respect to the coupling, i.e.\ $\beta(g)\partial_g$, while the Hamiltonian~(\ref{irevol}) will be second order and higher in $\pi$.  In our interpretation, these higher derivative terms correspond to the flow of the action out of the single-trace space.  
In the planar limit the HJ evolution can be written in terms of first derivatives, 
\begin{equation}
\partial_\ell W_{\rm IR}(\ell,\tilde\phi) = -H(\tilde\phi, i\delta W_{\rm IR}/\delta\tilde\phi)  \,,
\end{equation}
but it is nonlinear in first derivatives and so not of RG form.  

Finally, one can write an RG-like equation by expanding the HJ equation around a given solution $W_0$,
\begin{equation}
W_{\rm IR} = W_0 + \Gamma \,,
\end{equation}
Then at linear order
\begin{equation}
\left(\partial_\ell + i \frac{\delta H}{\delta \pi}(\tilde\phi, i\delta W_0/\delta\tilde\phi) \frac{\delta}{\delta\tilde\phi}\right) \Gamma(\ell,\tilde\phi) +O(\Gamma^2) =0\,,
\end{equation}
which does have the form of an \mbox{RG.}  In Ref.~\cite{de Boer:1999xf}, the zeroth order solution $W_0$ is taken to be the local part of the action, to some order in derivatives.  

We can illustrate this with the free scalar example of Sec.~3.1.2.  The leading local term~(\ref{freescalarultralocalIR}) gives
\begin{equation}
W_0 =  -\frac{\Delta_-}{2\ell^d}\int\frac{d^dk}{(2\pi)^d}\, \tilde\phi_{-k} \tilde\phi_k \,,
\end{equation}
and so
\begin{equation}
\ell \partial_\ell \Gamma(\ell,\tilde\phi) = -{\Delta_-} \int\frac{d^dk}{(2\pi)^d}\,  \tilde\phi_k \frac{\delta \Gamma}{\delta\tilde\phi_k} +O(\Gamma^2) \,.
\end{equation}
This is satisfied by the leading nonlocal term~(\ref{freescalarnlIR}),\footnote{As an aside, we note that the Wilsonian analysis~(\ref{wilir}) involved nonzero $\beta$-functions only for the unit and double-trace operators, whereas here there is only a single-trace $\beta$-function.}
\begin{equation}
\Gamma\approx {\rm const} \times  \ell^{-2\Delta_-}
 \int\frac{d^dk}{(2\pi)^d}\,  k^{2\nu}\tilde\phi_{-k} \tilde\phi_k  \,.
 \end{equation}
For the full nonlocal amplitude~(\ref{fir}) one must keep also the $\Gamma^2$ terms, which we have identified with the flow of multi-trace operators.  

This RG is in a non-Wilsonian scheme: the $\beta$-function depends on the particular solution $W_0$,  which depends on boundary conditions in the \mbox{IR.}  This is analogous to RG schemes where couplings are defined in terms of physical correlators at given momentum points, which depend to some extent on physics at all scales and on the particular vacuum the theory is in, whereas the Wilsonian couplings and evolution depend only on scales above the cutoff and are independent of the state.  This state-dependence is also emphasized in the Lorentzian discussion in Ref.~\cite{Lawrence:2006ze}.

Much of the HRG literature applies the formalism of Ref.~\cite{de Boer:1999xf}.  We emphasize that we are taking for our purposes a very specific definition of `Wilsonian,' and these other approaches may be quite useful in organizing the physics according to scale.  Other work that organizes the physics via a Wilsonian splitting but uses an IR renormalization scheme includes Ref.~\cite{Balasubramanian:1999jd} (the appearance of the Bessel function $K_\nu(kz)$ in the construction of the $\beta$ function indicates an IR scheme, because it is distinguished by finiteness at $z\to\infty$).  

Various papers take an approach more similar to ours, though we believe that much of our framework is new.  Ref.~\cite{Mansfield:1999kk}  analyzes properties of the radial Schrodinger flow.  Ref.~\cite{Lewandowski:2002rf}, in the context of Randall-Sundrum compactification, introduces a flow of the UV boundary action.  Ref.~\cite{Papadimitriou:2004ap} uses evolution of a radial slicing to study bulk divergences.  Eq.~(46) of that paper identifies a dilatation operator, but it involves only single-trace operators; we are not sure of the relation to our work.  Some properties of $W_{\rm UV}$ were considered in Refs.~\cite{Verlinde:1999xm, Ellwanger:2000pp}.
The recent work~\cite{Vecchi:2010dd} has some overlap with our general approach, in particular on the importance of multitrace operators, and with the calculations in Sec.~3.1.  Some relevant papers on the field theory side will be noted in the next section.

Recent work on critical behavior and transport in strongly coupled CFT's employ ideas closely related to ours.  Sec.~V of Ref.~\cite{Iqbal:2008by} gives an analysis of the flow between the horizon and the boundary that is similar in form to our discussion of domain wall flow.  Ref.~\cite{Bredberg:2010ky,FLR} also takes a Wilsonian approach to transport calculations, and Refs.~\cite{Faulkner:2010fh} discuss phase transitions driven by multi-trace flows.

\sect{The field theory side}

In discussions of the $1/N$ expansion single-trace actions are usually regarded as the norm, with multi-trace terms arising in more exotic applications. However, from the renormalization group point of view the reverse is true~\cite{Tseytlin:1999ii,Li:2000ec,Petkou:2002bb}.

 \begin{figure}
\begin{center}
\includegraphics[scale=.60]{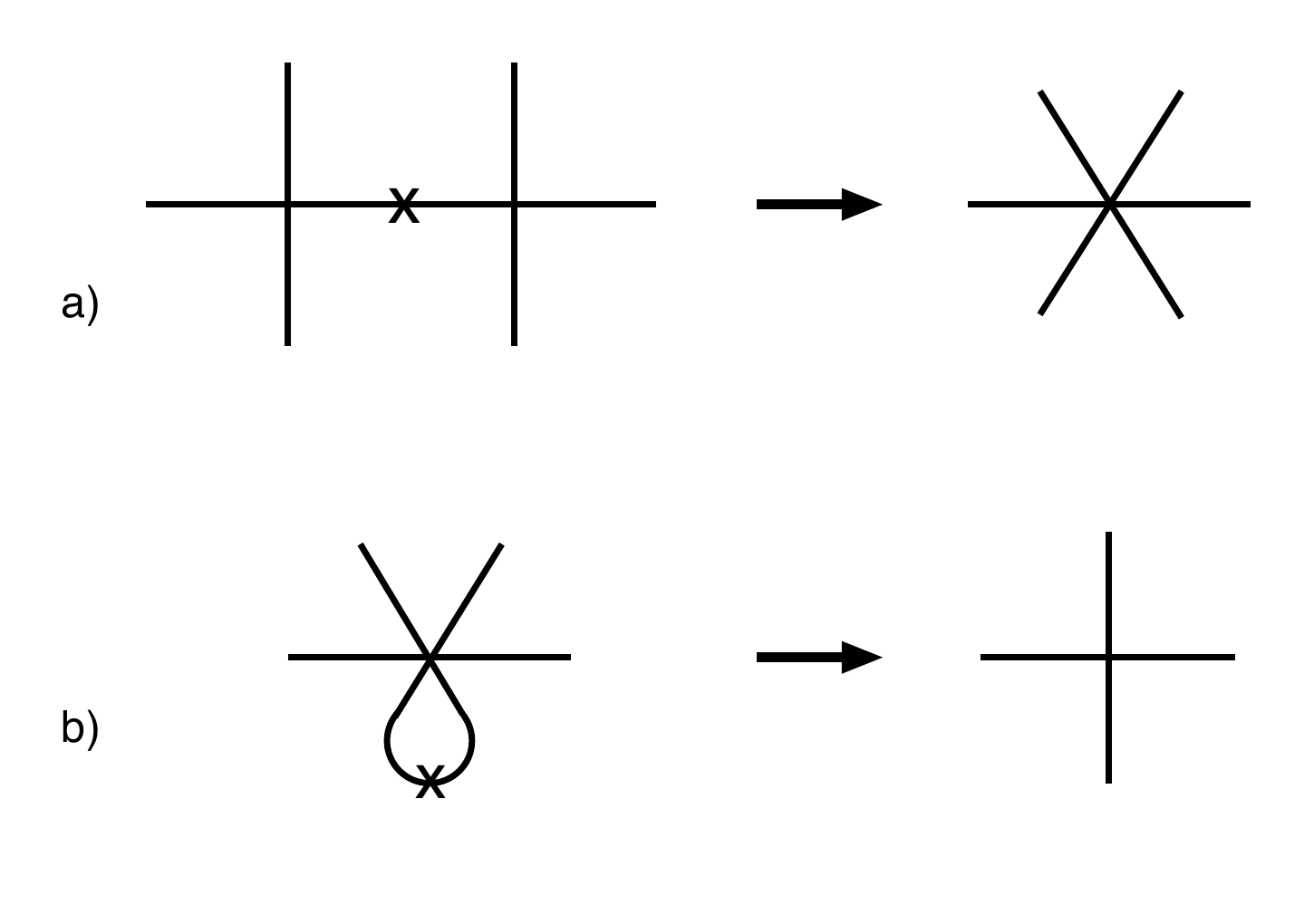} 
\caption{Terms in the Wilson RG.  a) Tree level term quadratic in the interaction.  b) One loop term linear in the interaction.}
\end{center}
\end{figure}
The schematic form of the Wilson RG~\cite{Wilson:1993dy}, in the form used in Ref.~\cite{Polchinski:1983gv},
\begin{equation}
\ell \partial_\ell S_{\rm int} = \dot\Delta \times \left( \frac{\delta S_{\rm int}}{\delta M}\frac{\delta S_{\rm int}}{\delta M} -  \frac{\delta^2 S_{\rm int}}{\delta M^2} \right)\,.
\end{equation}
Here $\dot\Delta$ represents the derivative of the propagator with respect to scale.  The first term arises when the integrated propagator connects two distinct vertices as in Fig.~2a, and the second arises when it connects a vertex to itself as in Fig.~2b.  The graphs provide useful intuition, but the derivation is nonperturbative.

For a matrix-valued theory, one can keep track of the effect on the multi-traces~\cite{Li:2000ec,Petkou:2002bb}. 
 \begin{figure}
\begin{center}
\includegraphics[scale=.50]{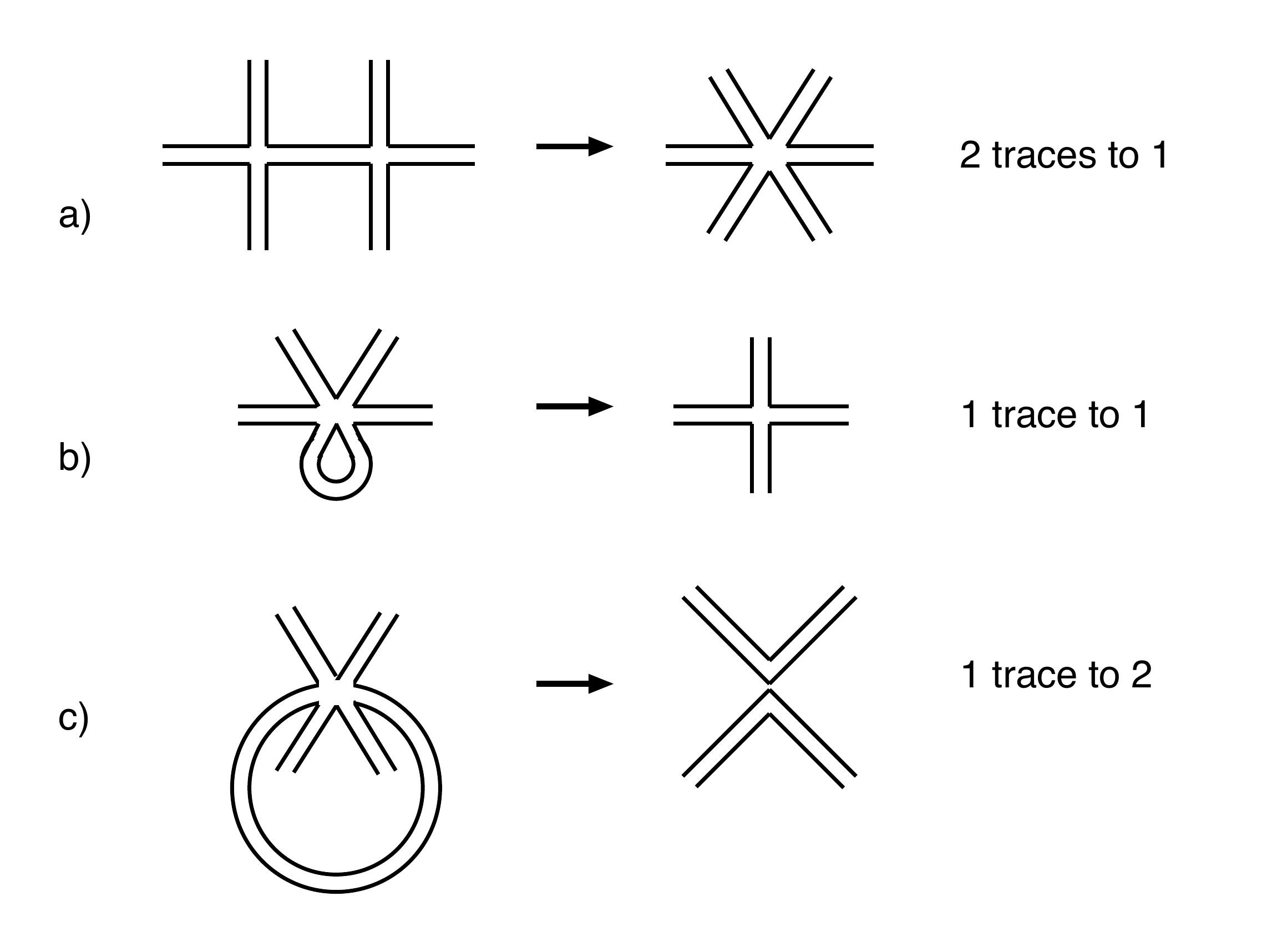} 
\caption{The Wilson RG in double line notation.}
\end{center}
\end{figure}
 It is convenient t     o rewrite the RG as
\begin{equation}
\ell \partial_\ell e^{-S_{\rm int}} = \dot\Delta \times  \frac{\delta^2}{\delta M^2}  e^{-S_{\rm int}}  \,.
\end{equation}
There are then three kinds of term, as in Fig.~3.   When the derivatives act on distinct traces they combine them into one: the action is of the form ${\cal O}\, \delta^2/\delta {\cal O}^2$.  When they act on a the same trace, they leave the number of traces fixed if they act on fields in adjacent cyclic order, a ${\cal O}\, \delta/\delta {\cal O}$ term, and they add a trace if the act on fields that are separated in the cyclic order, a ${\cal O}^2\, \delta/\delta {\cal O}$ term.  In all, the evolution is of the Schrodinger form~(\ref{evolve}) with
\begin{equation}
H_{\rm CFT} \sim {\cal O}\frac{\delta}{\delta {\cal O}} + N^{-2} {\cal O}\frac{\delta^2}{\delta {\cal O}^2} + {\cal O}^2\frac{\delta}{\delta {\cal O}} \,. \label{hcft}
\end{equation}
We have accounted a factors of $N$ from the propagator, the index loop, and the definition of $\cal O$.  All terms contribute in leading planar order, where $S$ is of the form $N^2 s({\cal O})$.

This cubic Hamiltonian does not at all resemble that of supergravity, but looks very much like string field theory, with $\delta/\delta{\cal O}$ destroying a string and $\cal O$ creating one.  This is in keeping with the remarks in Sec.~4.1.  It suggests that the Wilson RG should be related to to string theory quantized in a gauge where the AdS radius is identified with world-sheet time~\cite{Jevicki:1993rr, Fukuma:1993tp, Lifschytz:2000bj}.\footnote{Another approach connecting string field theory to the RG is Ref.~\cite{Brustein:1990wb}.  We note also the recent work~\cite{Lee:2009ij}; we are not sure how this relates to our approach, but it appears to differ in that the bulk fields will have the same quantum numbers as the boundary fields, and so will transform under the boundary gauge group.  Ref.~\cite{Akhmedov:2010sw} investigates the Wilson/holographic connection and argues for a truncation to the ultralocal sector in the planar limit.  We have noted below Eq.~(\ref{momex}) such a truncation with respect to derivatives acting on traced operators, but do not believe that one can also truncate with respect to derivatives acting on the separate fields within the trace.}$^,$\footnote{We should note that we have suppressed a general index sum in the Hamiltonian~(\ref{hcft}), and the term ${\cal O} = 1$ would look like a lower order interaction.  Also, the RG of the CFT acts not on the total action but on the interaction part, and the RG~(\ref{wilrg}) similarly acts on an action shifted by $S_0$; these shifts must be taken into account in comparing forms.}
 The nonpolynomial supergravity action is obtained from string field theory by integrating out the stringy fields, in parallel to the procedure described in Sec.~4.1.  
 
 The formal structure that we have developed in this paper will be useful only if we can give meaning to the relation~(\ref{ir}) between the cutoffs in the bulk and the gauge theory.  Of course, already on the gauge theory side it is intricate to construct a continuum cutoff that respects the gauge symmetry~\cite{Arnone:2006ie}; extended supersymmetry and dualities are also mangled in the Wilson framework.  As a final calculation in this paper, we would like to note that the Wilsonian holographic RG suggests a new approach to this problem.  
 
If we begin with the cutoff amplitude $\Psi_{\rm IR}(\ell,\tilde\phi)$ in the bulk, whose precise CFT dual is to be determined, we could try to use the Schr\"odinger evolution~(\ref{evolve}) 
\begin{eqnarray}
\Psi_{\rm IR}(\ell,\tilde\phi) &=& \int d\tilde\phi' \, G(\ell, \epsilon, \tilde\phi, \tilde\phi') \Psi_{\rm IR}(\epsilon,\tilde\phi')  \,,  \label{runir}
\end{eqnarray}
to express this in terms of $\Psi_{\rm IR}(\epsilon,\tilde\phi)$; for simplicity we are looking at the case that the backreaction is small.   
Taking the limit $\epsilon \to 0$, we reach the boundary and so the standard dictionary~\cite{Maldacena:1997re,Gubser:1998bc,Witten:1998qj} allows it to be translated into gauge theory variables.  Thus we identify a CFT calculation that produces the bulk amplitude with radial cutoff.  The problem is that we are running the evolution the wrong way: essentially, we are forming $e^{H\tau}$, which is not defined for a Hamiltonian unbounded above.  

Curiously, it seems that we can circumvent this by going to the Lorentzian theory, where an $i$ will appear in the evolution equation, giving an ordinary Schr\"odinger kernel.  (The construction is similar to that of bulk fields in AdS~\cite{holo}, which also seems to be essentially Lorentzian.)
For the quadratic scalar theory, one finds 
\begin{eqnarray}
\ln G(\ell, \epsilon, \tilde\phi, \tilde\phi') &=& \frac{ \tilde\phi^2(\Delta_+ x^{\Delta_+} - \Delta_- x^{\Delta_-}) - 2\tilde\phi\tilde\phi' x^d (\Delta_+ - \Delta_-)
+ \tilde\phi'^2 x^d (\Delta_+ x^{\Delta_-} - \Delta_- x^{\Delta_+}) }{-2i \ell^d( x^{\Delta_+} - x^{\Delta_-})} + O(k^2)\nonumber\\
&\approx& \frac{i  \tilde\phi^2 \Delta_+ }{2 \ell^d} - \frac{i \tilde\phi\tilde\phi' (\Delta_+ - \Delta_-)}{\ell^{\Delta^+} \epsilon^{\Delta_-}} - \frac{i \tilde\phi'^2 \Delta_- }{2\epsilon^d} \,, \label{rgback}
\end{eqnarray}
where $x =\ell/\epsilon$.  To interpret this, note that we can interpret the RHS of Eq.~(\ref{runir}) in
terms of a path integral for $\Psi_{\rm IR}$ with a free boundary condition and a boundary action from $G$.   The surface term in the saddle point equation for this path integral is
\begin{equation}
0 = - x^{-\Delta_+}{(\Delta_+ - \Delta_-) \phi(\ell)} + (\epsilon\partial_\epsilon -\Delta_-)\phi(\epsilon)
\end{equation}
This fixes the normalizable $\beta$ coefficient to the value $\ell^{-\Delta_+} \phi(\ell)$.  In other words, the effective path integral describes the alternate conformal quantization with a single trace perturbation (higher order terms dropped in the second line of Eq.~(\ref{rgback}) give rise also to a double-trace perturbation).

This applies to all fields, so leads to a strange result: the cutoff path integral corresponds to taking the alternate quantization for all operators; this is unphysical for $\Delta > (d-2)/2$, but that is to be expected in a regulated theory.  However, it is not clear in what sense this is a regulator.  A higher covariant derivative regulator would correspond to adding higher derivative single-trace terms, but of course this does not work in Yang-Mills theory.  The alternate quantization corresponds to adding double-trace terms, which begin at quartic order in the matrix fields and so do not affect the propagator directly.  In a sense we are treating the theory with alternate quantization as a starting point, and using its deformations to define the cutoff theory.\footnote{This is similar in flavor to Ref.~\cite{ArkaniHamed:1997mj}, which uses the ${\cal N}=4$ theory (with standard quantization) as a starting point for defining Wilsonian regulators of theories of less symmetry.  It may be that one should revise the Wilsonian strategy along these lines.}
We should note that the required step of running the RG backwards raises suspicions, though the theory with alternate quantization is actually an attractor for this flow in almost all directions, since most single-trace operators get negative dimensions. We leave further exploration of this idea for later work.

\sect{Final remarks}

We have identified structural parallels between the Wilson and holographic RG's.  The hope is that this will give a framework for making precise the connection~(\ref{ir}) between the cutoff theories, pursuing the various parallels we have noted.  The general direction of this work, if it succeeds, is to produce a more local version of holography, where the boundary is pulled into the bulk.  Of course, the radial slicing is not locally special, and so it should then be possible to discuss more general subregions of the bulk, with some sort of equivalence principle relating different slicings.  

Possibly a more modest goal is to extend the results of Ref.~\cite{Heemskerk:2009pn}, which found sufficient conditions for local physics to emerge in the bulk.  That approach of that paper used detailed properties of conformal blocks, and appears to rather technical to extend.  We believe that there should be a more flexible and physical way to organize the argument, perhaps using the tools we have developed here.


\section*{Acknowledgments}

We thank Abhay Ashtekay, Tom Faulkner, Steve Giddings, David Gross, Hong Liu, Mukund Rangamani, Eva Silverstein, Dam Son, and especially Don Marolf for discussions and for comments on the manuscript.  This work was supported in part by NSF grants PHY05-51164 and PHY07-57035. 


\end{document}